# Evaluating Factor Contributions for Sold Homes

Jason R. Bailey\*, W. Brent Lindquist, and Svetlozar T. Rachev

Department of Mathematics and Statistics, Texas Tech University, Lubbock, TX 79409-1042, USA

**Abstract.** We evaluate the contributions of ten intrinsic and extrinsic factors, including ESG (environmental, social, and governance) factors readily available from website data to individual home sale prices using a P-spline generalized additive model (GAM). We identify the relative significance of each factor by evaluating the change in adjusted $R^2$ value resulting from its removal from the model. We combine this with information from correlation matrices to identify the added predictive value of a factor. Based on data from 2022 through 2024 for three major U.S. cities, the GAM consistently achieved higher adjusted $R^2$ values across all cities (compared to a benchmark generalized linear model) and identified all factors as statistically significant at the 0.5% level. The tests revealed that living area and location (latitude, longitude) were the most significant factors; each independently adds predictive value. The ESG-related factors exhibited limited significance; two of them each adding independent predictive value. The elderly/disabled accessibility factor was much more significant in one retirement-oriented city. In all cities, the accessibility factor showed moderate correlation with one intrinsic factor. Despite the granularity of the ESG data, this study also represents a pivotal step toward integrating sustainability-related factors into predictive models for real estate valuation.

## 1. Introduction

Hedonic models are employed to capture the heterogeneous effects of intrinsic and extrinsic factors of residences and their location, respectively, on real estate prices. Using regression techniques, these models quantify the impact of each factor on the price of a house. The identification of relevant factors, the selection of the regression formulation, and the application of the model to real-world data constitute the three general steps in developing such models. In this paper, we consider a P-spline-based generalized additive model (GAM) for the valuation of completed sale transactions of homes based on intrinsic, extrinsic, and environmental, social, and governance (ESG) factors of the residences. Specifically, we develop a methodology for identifying the most significant contributory factors for home prices by evaluating the changes in adjusted $R^2$ values resulting from removing selected factors. Combining this with information from correlation matrices, we can identify significant factors that provide little added predictive value. We apply this methodology to home prices in three U.S. cities.

---

\* Corresponding author: Jason.R.Bailey@ttu.edu

The number of bedrooms and bathrooms, indoor and outdoor areas, and the categorization of the dwelling type (single-family, condominium, etc.) are commonly used and accepted intrinsic factors for real estate price models. More interesting are the variety of extrinsic factors considered. A well-known extrinsic factor is location (neighborhood desirability). Example location-related measures include postal codes and GPS coordinates, the latter providing more precise location granularity. Many publicly available geocoding websites provide the latitudinal and longitudinal coordinates of an estate, and such refinements have allowed for more extensive analyses. Eiling et al. (2019) used monthly housing returns for 9,831 zip codes across 178 U.S. Metropolitan Statistical Areas (MSAs) to quantify the idiosyncratic zip-code specific risk and systematic market risk within each MSA. Hill and Scholz (2018) established the superiority of a nonparametric spline surface based on GPS data over postal code proxy information. Helbich et al. (2013) examined the explanatory power of exposure to solar radiation on the pricing of owner-occupied flats in Vienna by employing airborne LIDAR maps. Olszewski et al. (2017) verified the significance of such factors as the distances to the nearest metro station, green space, and the city center. Cohen and Coughlin (2008) studied the effects of home proximity to airports.

Extrinsic macroeconomic factors have an effect on real estate prices. In their study, Olszewski et al. (2017) also analyzed the effects of housing policy on prices. Belke and Keil (2017) investigated several macroeconomic factors, including the per-capita number of newly constructed apartments, the per-capita number of real estate market transactions, the unemployment rate, the purchasing power index of the area, and the number of hospitals.

The components of ESG represent the sustainability factors of a property. The risk of a natural disaster, the installation of renewable energy systems, and resiliency to global warming are examples of environmental factors. Construction worker labor standards, homeowner satisfaction, and noise pollution are examples of social factors. Regulatory compliance with standards set at all governmental levels, overall transparency, and legal issues related to property owner practices are examples of governance factors.

Lauper et al. (2013) analyzed the green home acquisition and installation process from the point of view of a homebuilder. Social norms and policies have been shown to not only heighten consumer spending and interest on environmentally friendly appliances but also significantly impact the energy-relevant decisions made during homebuilding (Reposa 2009, Palm 2017, Rakha 2018). Ma et al. (2019) analyzed the impact of governmental policymaking processes on the adoption of residential green energy additions and construction. Specifically, they noted that stringent governmental policies on residential green energy subsidies can have an adverse effect on household installations.

Under global climate changes, environmental factors (flood risk, pollution, wildfires, number of extreme temperature days, etc.) can be expected to play a role in homebuyer decisions and, as a consequence, real estate pricing. Lavaine (2019) found that the closure of a toxic site leading to a decrease in atmospheric $SO_2$ levels was associated with an increase in the average house price but a decrease in average price of flats. Quantitative environmental indices have been developed to provide guidance to consumers in assessing house prices. Mahanama et al. (2021) formulated an index to measure the level of future systemic risk caused by natural disasters. The study by Contat et al. (2023) confirmed that the risks of wildfire and flooding correlated inversely with home prices, as the higher risks resulted in discounts on said prices.

As hedonic models aim to estimate the contributory value of each external or internal factor, the decomposition allows for the appropriate use of generalized additive, logarithmic, or linear models to identify the contributive power of each factor. Pace (1998) was one of the earliest to use a GAM in the context of real estate pricing and demonstrated that GAMs could outperform more unsophisticated polynomial and parametric models. Owusu-Ansah (2011) presented a review of semi-parametric, parametric, and non-parametric models. Silver (2016) proposed a hedonic regression pricing methodology. Colonnello et al. (2021) considered a linear hedonic model for housing yield (rent-to-price ratio) and incorporated a relatively large number of demographic, local economic, and extrinsic factors. Brunauer et al. (2013) used a four-level hierarchical additive regression model to quantify the contribution of each level of geographic detail to housing prices. Bárcena et al. (2013) employed a geographically weighted and semi-parametric hedonic model to create an index of housing prices in Bilbao, Spain, over the time period before and after the Great Recession. Bax and Chasomeris (2019) used a generalized linear model (GLM) to measure apartment rent prices from a set of statistically significant factors.

Doszyń and Gnat (2017) used predictive and studentized residuals of a properly specified linear model to regress the price per square meter of plots of land to six factors. However, many models are not properly specified and correctly applied, which can result in models of poor quality, an issue especially prevalent with linear regression models. A series of simulations involving varying levels of price disturbances in linear, multiple regression models found that even minor disturbances meaningfully reduced $R^2$ values (Kokot and Gnat 2019). Linear models can only be expected to perform adequately in well-developed and well-functioning real estate markets whose influencing factors on real estate valuations exhibited strong and linear relationships.

Bailey et al. (2022) used intrinsic and extrinsic factors (including some ESG factors) in GAM and GLM models to analyze the variance in the logarithm of the expected sales price of homes. When ESG factors easily accessible from real estate vendor websites were included, minor improvements were observed in the adjusted $R^2$ values of the model. A further analysis using these ESG factors and new home constructions to estimate the average annual home prices of eight U.S. cities over two decades found that the ESG factors had city-dependent significance in predictive power (Bailey et. al. 2024). In both studies, GAMs were found to significantly outperform GLMs.

We note that intrinsic and extrinsic factors have been shown to have varying impact on home valuations within the distribution of the local housing prices. Analyses of home sales found that segmentation by price quantile was vital in the assessment of the impact of several input factors (Zeitz et al. 2007, 2008; Lian and Wang 2020).

## 2. Materials and Methods

### 2.1 Price and Factor Data

Our data set[1] was based on completed sale transactions of homes within the 36-month period preceding the end of 2024 (see Appendix A for the collection process). Data were assembled for the cities of Denver, CO. (DEN), Jacksonville, FL. (JAX), and Phoenix, AX (PHX). The data set consisted of dwelling price and ten factors, seven of which are intrisic: living area (SqFt), lot size (Lot), the number of bedrooms (Beds), the number of bathrooms (Baths), the year during which the construction of the dwelling was completed (Year), whether the home was green-rated (Green), and whether the home was considered accessible to the elderly and disabled (Access); and three extrinsic location factors: latitude (Lat), longitude (Long), and whether the home was along a waterfront (Water),. Of these, Green, Access and Water are ESG factors. We note that the data are restricted to home sales within city limits. Due to the heavy-tailed nature of dwelling prices, we used $\log_{10}$(Price) to express dwelling price (log-price). The seven non-binary factors (SqFt, Lot, Beds, Baths, Lat, Long, and Year) are normalized for this study by computing the sample mean and standard deviation for each factor and converting each data value to a z-score (Bailey et. al. 2024).

---

[1] Price and factor data were obtained from Redfin.com. Data obtained by specification of the city and the entries for "All filters". Appendix A provides the specific filter values.

## 2.2 The Generalized Additive and Linear Models

A GAM relates a univariate response variable $Y$ to a set of predictive (intrinsic and extrinsic) factors $x_j$, $j = 1, \ldots, m$ (Hastie and Tibshirani 1990). It relates the expected value $\mu = E[Y]$ to the factors through functional dependencies,

$$g(\mu) = \beta_0 + f_1(x_1) + f_2(x_2) + \cdots + f_m(x_m). \tag{1}$$

It is assumed that $Y \sim EF(\mu, \theta)$, where $EF(\mu, \theta)$ denotes the exponential family of distributions with mean $\mu$ and scale parameter $\theta$. The link function $g(\cdot)$ relates conditional expectations of $Y$ to the factors through

$$\mu = g^{-1}\big(\beta_0 + f_1(x_1) + f_2(x_2) + \cdots + f_m(x_m)\big). \tag{2}$$

We used the identify function for $g(\cdot)$ and P-splines (Eilers and Marx 1996) for the functions $f_j(\cdot)$, which minimize the penalized sum of squares

$$\sum_{i=1}^{N} \left( Y_i - \sum_{j=1}^{m} f_j(x_{ji}) \right)^2 + \sum_{j=1}^{m} \lambda_j \int f_j''(z)^2 dz, \tag{3}$$

where $Y_i$, $i = 1, \ldots, N$, and $x_{ji}$, $j = 1, \ldots, m$, $i = 1, \ldots, N$, are $N$ data observations. The weight given to the smoothness of function $f_j(\cdot)$ is determined by the tuning parameter $\lambda_j > 0$. The values $x_{ji}$, $i = 1, \ldots, N$, are referred to as the knots of the function $f_j(\cdot)$.

As a benchmark, we compared the results obtained from the GAM to those obtained from a standard GLM of the form

$$g\big(E_Y(Y|X)\big) = \beta_0 + \beta_1 x_1 + \cdots + \beta_m x_m \equiv X\beta. \tag{4}$$

In (4), $X$ is a $N \times (m+1)$ matrix, $Y = [Y_1, ..., Y_N]^T$ is the column vector of values of the response variable, $x_j = [x_{j1}, ..., x_{jN}]^T$ is the column vector of values for factor $x_j$, and $\beta = [\beta_0, \beta_1 ..., \beta_m]^T$ is the column vector of unknown parameters. The first column of $X$ is a vector of ones and the remaining columns correspond to the factor column vectors. As the identity function was used for $g(\cdot)$, (4) becomes a pure linear model.

## 3. Results

The models were run using the R *gam* package and the *lm* function (Hastie 2023). The *p*-values associated with the factors for each of the three cities as fit by the models are shown in Table 1. The number of factors with *p*-values below 0.01 was larger for the GAMs than the GLMs in all three cities. Additionally, every factor was found to be significant at a *p*-value threshold of 0.01 for all three GAMs, whereas each GLM had at least one factor *p*-value in excess of 0.05. Six to ten percentage point increases occurred in the adjusted $R^2$ for all three cities under the GAM.

**Table 1.** Significance (*p*-value) of the factors in the GLM and GAM fits.

| Factor | DEN | JAX | PHX | DEN | JAX | PHX |
|---|---|---|---|---|---|---|
| | | GLM | | | GAM | |
| SqFt | *** | *** | *** | *** | *** | *** |
| Lot | *** | 0.104 | *** | *** | *** | *** |
| Beds | *** | *** | *** | *** | *** | *** |
| Baths | *** | *** | *** | *** | *** | *** |
| Lat | *** | *** | *** | *** | *** | *** |
| Long | *** | *** | *** | *** | *** | *** |
| Year | *** | *** | *** | *** | *** | *** |
| Water[a] | 0.086 | *** | *** | *** | *** | *** |
| Green | *** | 0.044 | 0.282 | *** | 0.003 | *** |
| Access | 0.010 | 0.005 | *** | *** | *** | *** |
| Adj. $R^2$ | 0.666 | 0.623 | 0.694 | 0.761 | 0.687 | 0.769 |

\*\*\* Indicates *p*-value < 0.001.　　[a] ESG factors highlighted in green font.

In order to assess the contribution of each factor to the predictive power of the models, each GAM was rerun with each computation having one of the ten factors excluded. We computed $\Delta R^2$ defined as the GAM baseline adjusted $R^2$ from Table 1 minus the adjusted $R^2$ from rerunning the model with one factor dropped. Hence a positive value of $\Delta R^2$ corresponds to a decrease in the adjusted $R^2$ when the factor is removed. The resulting changes are shown in Table 2. For each city, the factors have been listed in order of the magnitude of $\Delta R^2$. The ESG factors are highlighted in green font.

Defining the significance of a factor by its $\Delta R^2$ impact, for all three cities the home square footage, and its longitude and latitude are the three most significant pricing factors. The latter two factors reflect the well-known importance of location in housing price, while square footage reflects the costs of construction, maintenance and luxury. Whether longitude or latitude is more significant reflects the north vs. south, east vs. west variation in affluence in a city. What begins to differentiate the cities are the 4'th and 5'th most significant factors. While the year of construction (reflecting age as well as the change in society's housing "tastes" and preferences with time) constitutes one of these two factors for each city, the other factor varies by city, with number of bathrooms having stronger significance than year of construction for Denver, while lot size is competitive with year of construction for Phoenix. For Phoenix, lot size is just as significant a factor as year of construction. Interestingly the accessibility ESG factor is the 5'th factor of significance for Jacksonville. Arizona and Florida (home to Phoenix and Jacksonville, respectively) are both well-known retirement states. While 14.6% of the population of Jacksonville and 11.9% of Phoenix are older adult (65+), that difference alone does not explain why accessibility is more significant in Jacksonville ($\Delta R^2 = 0.014$) than in Phoenix ($\Delta R^2 = 0.003$). In fact, Denver has an older-adult population percentage of 12.3%, slightly larger than Phoenix, but its accessibility factor has no significant effect on the explained variance in the GAM fit.

**Table 2.** Value of $\Delta R^2$ for the GAM fits with the stated factor removed.

| DEN | | JAX | | PHX | |
|---|---|---|---|---|---|
| SqFt | 0.043 | SqFt | 0.082 | Long | 0.078 |
| Long | 0.040 | Long | 0.051 | SqFt | 0.036 |
| Lat | 0.023 | Lat | 0.027 | Lat | 0.027 |
| | | | | | |
| Baths | 0.023 | Year | 0.022 | Year | 0.023 |
| Year | 0.010 | Access | 0.014 | Lot | 0.022 |
| | | | | | |
| Lot | 0.002 | Baths | 0.006 | Baths | 0.005 |
| Beds | 0.001 | Water | 0.005 | Beds | 0.004 |
| Water[a] | *** | Lot | 0.004 | Access | 0.003 |
| Green | *** | Beds | 0.003 | Water | *** |
| Access | *** | Green | 0.001 | Green | *** |

\*\*\* Indicates change in adjusted $R^2$ of $< 0.001$.    [a] ESG factors highlighted in green font.

The five least significant factors are populated (with the noted exception of accessibility for Jacksonville) by the ESG factors, as well as number of bedrooms, number of bathrooms (with the noted exception of Denver) and lot size (with the noted exception of Phoenix). While the number of bedrooms and bathrooms always appear in real estate listings, reflecting the importance of family size to a potential buyer, their impact on house pricing is "not all that significant". Where lot size is not significant as a factor, city density may preclude substantial variation in lot sizes. Jacksonville was the only city with a measurable $\Delta R^2$ for all three ESG factors. Its relative significance of the waterfront factor (compared to lot size and number of bedrooms) may reflect Jacksonville's acreage bordering the Atlantic Ocean and the St John's River which meanders through the city.

With $\Delta R^2$ used as a measure of factor significance, we considered factor–factor correlations to measure added predictive value. For each city, we computed the correlation matrix $\boldsymbol{R} = [r_{ij}]$, $i = 1, \ldots, 10, j = 1, \ldots, 10$, where

$$r_{ij} = \frac{\text{cov}(\boldsymbol{x}_i, \boldsymbol{x}_j)}{\sigma_{x_i} \sigma_{x_j}} \tag{5}$$

is the Pearson correlation coefficient between factor observations $x_i$ and $x_j$ with $\text{cov}(x_i, x_j)$ denoting the sample covariance; and $\sigma_{x_i}$ and $\sigma_{x_j}$ the sample standard deviations. The correlation matrices for each city are displayed in Tables B1 – B3 in Appendix B. In assessing the Pearson correlation values, we adopt the qualitative descriptions of the strength of association (of a pair of variables) summarized in Table 3.

**Table 3.** Strength of association descriptions

| $|r_{i,j}|$ | Description |
|---|---|
| [0.00, 0.10) | Negligible |
| [0.10, 0.30) | Weak |
| [0.30, 0.50) | Moderate |
| [0.50, 0.70) | Strong |
| [0.70, 1.00] | Very Strong |

If the strength of association between factors $i$ and $j$ is strong ($|r_{i,j}| \geq 0.50$), then used together the two factors are adding little predictive value and one could be dropped from the model. If their associate is weak ($|r_{i,j}| < 0.30$) both factors are needed. In the case of moderate association ($|r_{i,j}| \in [0.30, 0.50)$), either both can be kept, or a different combination of the two factors should be considered.

Tables B1 – B3 provide the correlation values $r_{\text{SqFt,Lat}}$, $r_{\text{SqFt,Long}}$, $r_{\text{Lat,Long}}$ for the three most significant factors. All values are weak, with the exception of $r_{\text{Lat,Long}} = 0.33$ for Denver, which is moderate. Thus, these three factors each provide added predictive value.

The relevant correlation values $r_{i,\text{SqFt}}$, $r_{i,\text{Long}}$, $r_{i,\text{Lat}}$ and $r_{5,4}$ for the $i$ = 4th and 5th significant factors are given in data rows 4 and 5 of Tables B1 – B3. For Denver, $r_{\text{Baths,SqFt}}$ is very strong, indicating that the number of bathrooms is not adding a great deal of predictive power to the GAM model. This is reinforced by the observation that $r_{\text{Year,Baths}}$ has a moderate correlation. For Jacksonville, the correlations shown in Table 5 are all negligible or weak, indicating that both year of construction and accessibility add predictive value to the GAM. For Phoenix, both $r_{\text{Year,SqFt}}$ and $r_{\text{Lot,SqFt}}$ have moderate values, perhaps indicative of a general construction increase in house size over time combined with movement out of a denser city center.

Finally, we consider the correlation values (last five data rows of Tables B1 – B3) corresponding to the least significant factors. The data indicate a very strong value for $r_{\text{Baths,SqFt}}$ and strong values of $r_{\text{Beds,SqFt}}$ and $r_{\text{Baths,Beds}}$. The strong correlation values $r_{\text{Baths,SqFt}}$ and $r_{\text{Beds,SqFt}}$ are an obvious reflection of construction, although it is interesting that $r_{\text{Baths,SqFt}} > r_{\text{Beds,SqFt}}$. The strong correlation values of $r_{\text{Baths,Beds}}$ reflects family dynamics. Since numbers of bedrooms and bathrooms are not generating significant $\Delta R^2$ values (with the exception of Baths for Denver), their inclusion in the GAM model is not adding significant predictive value.

For Denver, all other correlations are negligible or weak except $r_{\text{Access,SqFt}}$ and $r_{\text{Access,Bath}}$. The magnitudes are moderate; interestingly the signs of both correlations are negative. For Jacksonville, all other correlations are negligible or weak except for $r_{\text{Baths,Year}}$ and $r_{\text{Baths,Access}}$, which are moderate in magnitude. For Phoenix, all other correlations are negligible or weak except for $r_{\text{Baths,Year}}$, $r_{\text{Baths,Lot}}$ and $r_{\text{Access,Baths}}$. For all three cities, $r_{\text{Access,Baths}}$ is negative. This negative correlation may correspond to the fact that older-adult homes often correspond to retired, "empty nesters" who have downsized their living quarters.

## 4. Discussion

The significance and correlation tests revealed that living area and latitude-longitude location exerted the strongest impact of the explained variance of the GAM and each independently adds predictive value. The numbers of bedrooms and bathrooms correlate strongly with living area, indicating that these two factors, commonly included in hedonic housing models, add little predictive value. The year of construction completion emerged as the 4th or 5th most significant factor in each city, always having moderate correlation with the number of bathrooms (and additional moderate correlation with square footage and longitude in the case of Denver).

Of the ESG factors, the waterfront and green energy binary factors had low significance but did add predictive value independent of the other factors. The ESG factors had greater significance for Jacksonville than for the other two cities, with the accessibility factor being the 5th most significant. For each city, the accessibility factor had had moderate, negative correlation with the number of bathrooms.

Except for Phoenix, lot size provided low significance but did add predictive value. For Phoenix, lot size was competitive in significance with year of construction and was moderately correlated with living area and number of bathrooms. The lot size correlation with living area seems natural; the fact that lot size had moderate correlation with number of bathrooms undoubtedly just reflects the very strong correlation between SqFt and Baths.

One limitation of this study is that all three qualitative ESG factors do not contain any further specificity than was available in the dataset. For example, a home was coded as either not having green energy (0) or having green energy (1). However, different green energy units can have varying efficiency ratings. We would expect efficiency levels (such as solar panel quality) to factor into a home's valuation. Suppose that a solar panel factor was stratified based on effectiveness; for example, none (0), low (1), medium (2), and high (3). While this would represent an improvement in data refinement, it presupposes a stratification into "equidistant" intervals. However, a homebuyer may view the difference between low and medium quality panels to be more significant than that between medium and high-quality panels. We speculate that more precise information, such as age and/or wattage of the panels would further strengthen the model's predictive power.

Our prior work has considered the presence of central air conditioning as an ESG factor. Although the option existed to filter by central air conditioning, no homes were specifically identified as such in our Redfin data set for these three cities. Given that prior literature (see the Introduction) has identified the significance of central air conditioning on home pricing, we were unfortunately unable to estimate its contribution to the adjusted $R^2$ values.

**Data Availability**. The study's data is available upon request to the corresponding author.
**Code Availability**. The study's source code is available upon request to the corresponding author.
**Conflicts of Interest.** The authors declare no conflict of interest.

# Appendix A

Data were downloaded via a Python script accessing RedFin's application programming interface (API). The data-filter values specified in the script are provided in Table A1. As RedFin limits API requests to ten thousand homes, the script collected the data in "bundles" of ascending price until data for all homes was obtained.

**Table A1.** Filter values used for the RedFin data.

| Filter | Input | Filter | Input |
|---|---|---|---|
| Status | Sold, Last 36 Months | | |
| Price Range | MIN: $100k, MAX: $10M | | |
| Number of Bedrooms | 1+ | | |
| Number of Bathrooms | 1+ | | |
| Home Type | House | | |
| | **More Filters** | | |
| Square Feet | MIN: 750, MAX: NS[1] | Stories | MIN: NS, MAX: NS |
| Lot Size | MIN: 1000, MAX: NS | Year Built | MIN: NS, MAX: NS |
| Garage Spots | NS | Pool Type | NS |
| Exclude 55+ Communities | NS | Basement | NS |
| Air Conditioning | NS | Washer/Dryer Hookup | NS |
| Fireplace | NS | Elevator | NS |
| Primary Bedroom on Main Floor | NS | Pets Allowed | NS |
| Guest House | NS | Has a View | NS |
| Waterfront | ESG[2] | Fixer-Upper | NS |
| Green Home | ESG | Accessible Home | ESG |

[1] NS = Not Specified    [2] "Yes" when filtering for those houses and "NS" otherwise

# Appendix B

**Table B1.** Pearson correlation coefficients, $r_{\text{row factor, column factor}}$, for Denver

| | SqFt | Long | Lat | Baths | Year | Lot | Beds | Water | Green | Access |
|---|---|---|---|---|---|---|---|---|---|---|
| SqFt | 1.00 | | | | | | | | | |
| Long | 0.11 | 1.00 | | | | | | | | |
| Lat | -0.15 | 0.33 | 1.00 | | | | | | | |
| Baths | 0.83 | 0.18 | -0.09 | 1.00 | | | | | | |
| Year | 0.37 | 0.42 | 0.09 | 0.44 | 1.00 | | | | | |
| Lot | 0.26 | -0.13 | -0.25 | 0.14 | 0.04 | 1.00 | | | | |
| Beds | 0.63 | 0.05 | -0.11 | 0.63 | 0.26 | 0.20 | 1.00 | | | |

|       | SqFt | Long | Lat | Year | Access | Baths | Water | Lot | Beds | Green |
|-------|------|------|-----|------|--------|-------|-------|-----|------|-------|
| Water | 0.08 | -0.04 | -0.05 | 0.05 | 0.05 | 0.05 | 0.02 | 1.00 | | |
| Green | 0.05 | 0.00 | 0.00 | 0.05 | 0.03 | 0.01 | 0.04 | 0.00 | 1.00 | |
| Access | -0.37 | -0.25 | -0.08 | -0.46 | -0.28 | 0.10 | -0.21 | 0.00 | 0.00 | 1.00 |

Colors indicate strength of association: white – negligible, weak; green – moderate; light red – strong; dark red – very strong.

**Table B2.** Pearson correlation coefficients, $r_{\text{row factor, column factor}}$, for Jacksonville

|       | SqFt | Long | Lat | Year | Access | Baths | Water | Lot | Beds | Green |
|-------|------|------|-----|------|--------|-------|-------|-----|------|-------|
| SqFt  | 1.00 | | | | | | | | | |
| Long  | 0.18 | 1.00 | | | | | | | | |
| Lat   | -0.11 | 0.00 | 1.00 | | | | | | | |
| Year  | 0.25 | 0.11 | -0.04 | 1.00 | | | | | | |
| Access | -0.27 | -0.05 | 0.01 | -0.07 | 1.00 | | | | | |
| Baths | 0.81 | 0.17 | -0.11 | 0.34 | -0.31 | 1.00 | | | | |
| Water | 0.28 | 0.12 | -0.01 | 0.11 | -0.02 | 0.22 | 1.00 | | | |
| Lot   | 0.03 | 0.00 | 0.01 | -0.05 | -0.04 | 0.19 | 0.23 | 1.00 | | |
| Beds  | 0.61 | 0.06 | -0.01 | 0.25 | -0.17 | 0.60 | 0.13 | 0.13 | 1.00 | |
| Green | 0.03 | 0.01 | 0.00 | 0.02 | -0.01 | 0.03 | 0.04 | 0.02 | 0.03 | 1.00 |

Colors indicate strength of association: white – negligible, weak; green – moderate; light red – strong; dark red – very strong.

**Table B3.** Pearson correlation coefficients, $r_{\text{row factor, column factor}}$, for Phoenix

|       | Long | SqFt | Lat | Year | Lot | Baths | Beds | Access | Water | Green |
|-------|------|------|-----|------|-----|-------|------|--------|-------|-------|
| Long  | 1.00 | | | | | | | | | |
| SqFt  | 0.20 | 1.00 | | | | | | | | |
| Lat   | 0.13 | 0.13 | 1.00 | | | | | | | |
| Year  | -0.14 | 0.32 | 0.11 | 1.00 | | | | | | |
| Lot   | 0.17 | 0.46 | 0.16 | -0.02 | 1.00 | | | | | |
| Baths | 0.15 | 0.81 | 0.08 | 0.33 | 0.35 | 1.00 | | | | |
| Beds  | 0.02 | 0.64 | 0.03 | 0.21 | 0.23 | 0.61 | 1.00 | | | |
| Access | -0.22 | -0.37 | -0.14 | -0.26 | -0.07 | -0.33 | -0.26 | 1.00 | | |
| Water | 0.02 | 0.02 | -0.03 | 0.01 | 0.01 | 0.02 | 0.00 | -0.02 | 1.00 | |
| Green | 0.04 | 0.07 | 0.02 | 0.02 | 0.03 | 0.07 | 0.06 | -0.04 | 0.03 | 1.00 |

Colors indicate strength of association: white – negligible, weak; green – moderate; light red – strong; dark red – very strong.